\documentclass[seceq]{ptptex}
\newcommand{\bd}[1]{\mbox{\boldmath$#1$}}
\usepackage{graphicx}

\title{Rotating Spirals without Phase Singularity in Reaction-Diffusion Systems}

\author{Yoshiki {\sc Kuramoto} and Shin-ichiro {\sc Shima}}  

\inst{Department of Physics, Graduate School of Sciences, 
      Kyoto University, Kyoto 606-8502, Japan}

\recdate{ }

\abst{Rotating spiral waves without phase singularity are found to arise in a 
certain class of three-component reaction-diffusion systems of biological
relevance. It is argued that this phenomenon is universal
when some chemical components involved are diffusion-free. 
Some more detailed mathematical and numerical analyses are carried out
on a complex Ginzburg-Landau equation with non-local coupling
to which the original system is reduced 
close to a codimension-two parameter set.   
}

\begin{document}

\maketitle

\section{Introduction}
Rotating spiral waves represent a most universal class of patterns
observed in extended fields composed of excitable
or self-oscillatory local elements. Recent experimental and theoretical studies
on rotating spirals have focused exclusively on their complex behavior such as
core meandering in 2D\cite{rf:1}, control of the spiral dynamics using photo-sensitive
BZ reaction\cite{rf:2}, and the topology and dynamics of the singular 
filaments 
of 3D scroll waves\cite{rf:3}. In the present article, we argue that there must be 
yet another problem of fundamental importance associated with 
the spiral dynamics\cite{rf:4}.
This is the occurrence of spatial discontinuity
near the core, which also leads to the loss of phase singularity. 
In reaction-diffusion
systems, occurrence of spatial
discontinuity itself is not surprising when the system involves some components without diffusion. Thus, our main concern will be under what general conditions
such behavior arises, why the core is the weakest region 
regarding the loss of spatial continuity, and how singular is the 
structure 
of the core
after such breakdown.

In \S 2, we introduce a simple class of reaction-diffusion systems
which can exhibit 2D spiral waves without phase singularity.
The specific system studied is a field of continuously distributed 
FitzHugh-Nagumo
oscillators without direct diffusive coupling supplemented with an extra 
chemical component which is the only
diffusive component thus mediating the coupling among the local oscillators.
Our numerical simulation suggests that the coupling strength $K$ between the
diffusive component and the local 
oscillators is the crucial parameter, and that
weaker coupling implies core anomaly. 
In order to proceed to more detailed analysis of the core anomaly, we apply
in \S 3 a method of reduction to our reaction-diffusion system 
near the Hopf bifurcation. It is known that such reduction generally leads to
the complex Ginzburg-Landau (CGL) equation which by no means exhibits
the type of core anomaly of our concern.
However
, there is an exceptional case in which $K$ is comparable in magnitude 
with the 
bifurcation parameter, and in that particular case
the diffusion coupling in CGL must be replaced with
a non-local coupling. We take advantage of this fact for the purpose of detailed
analysis of the core anomaly using this simple and universal equation.
In \S 4, some mathematical and numerical study 
of our non-local CGL will be presented.
First, it is almost trivially concluded that 
there is a critical 
condition across which steadily rotating waves
with vanishing amplitude at the center of rotation becomes impossible.
We also present some numerical results on the core anomaly
exhibited by the non-local CGL.
Section 5 is devoted to a speculative argument concerning possible statistical
natures of the turbulent fluctuations inside the core. The point of our
argument is to note the fact that the system has no characteristic length
scale inside the core so that the turbulent fluctuations there should 
be scale-free or their statistics should be characterized by some scaling laws.
A few concluding remarks will be given in \S 6.

\section{Loss of spatial continuity in the spiral core}
In some previous works on spatio-temporal chaos in self-oscillatory 
media\cite{rf:5,rf:6,rf:8},
a particular class of three-component 
reaction-diffusion systems of the following form was considered:
\begin{eqnarray}
\dot{X}&=&f(X,Y)+KB, \\
\dot{Y}&=&g(X,Y), \\
\tau\dot{B}&=&-B+D\nabla B+X, 
\end{eqnarray}
where the set of equations $(\dot{X},\dot{Y})=(f,g)$ represents a 
local limit-cycle 
oscillator. 
The above model, possibly with various generalizations, may serve as a suitable
model for a large assembly of oscillatory units (represented by the first 
two equations) without direct mutual coupling supplemented with an extra 
component (represented by the last equation) which behaves as a coupling 
agent among the local 
oscillators. Our system, possibly with various modifications, 
bears some resemblance to biological assemblies of oscillatory and excitable
cells such as yeast cells under glycolysis and slime mold amoebae.
It has also some similarity with the recently developed experimental 
system of the Belousov-Zhabotinsky reaction dispersed in water-in-oil AOT
microemulsion\cite{rf:7}.
For the sake of convenience, parameter $\tau$
has been inserted in the last equation 
to indicate the time scale of the diffusive
component $B$. When $\tau$ is sufficiently small, adiabatic elimination of $B$
, which can be done explicitly because the equation for $B$ is linear, 
leads to a two-component
non-locally coupled system of the form
\begin{eqnarray}
\dot{X}(\bd{x},t)&=&f(X,Y)+K\int G(|\bd{x}-\bd{x'}|)X(\bd{x'},t)d\bd{x}', \\
\dot{Y}(\bd{x},t)&=&g(X,Y).
\end{eqnarray}
The spatial extension of our system is assumed to be sufficiently large,
so that in two-dimensional systems to which our discussion below will 
exclusively be
confined, $G(r)$ is given by a modified Bessel function $K_0(r/D^{1/2})\cdot(2\pi D)^{-1}$.
Note that the effective coupling range is of the order of $D^{1/2}$.

Diffusion-coupling approximation of the above non-local coupling is valid
if the pattern obtained from Eqs.~(2.4) and (2.5) is such that the characteristic wavelength of $X$,
denoted by $l_p$,
is sufficiently longer than the coupling range, i.e., if
\begin{equation}
l_p\gg D^{1/2}.
\end{equation} 
Under the above condition, our system behaves similarly to a two-component
reaction-diffusion systems with effective diffusion constant
of the order of $KD$, thus arousing no particular interest 
associated
with the dynamics peculiar to non-locality in coupling.
Since the result of the diffusion-coupling approximation made on
Eq.~(2.4) implies that
the characteristic wavelength of the pattern scales like
\begin{equation}
l_p\sim (KD)^{1/2},
\end{equation}
this approximation is consistent only if $(KD)^{1/2}\gg D^{1/2}$ or
only if the coupling strength $K$ is sufficiently large. 
Thus, our main concern below is how our system behaves as $K$ is made
smaller by which something peculiar to non-locally coupled systems is
expected to emerge. Among others,  
this article is  particularly concerned with the dynamics
of rotating spiral waves.

As a simple model for the local oscillators, let us choose the 
FitzHugh-Nagumo model
\begin{equation}
f=\epsilon^{-1}\{(X-X^3)-Y\}, \quad g=aX+b
\end{equation}
under an oscillatory condition $a=1.0$, $b=0.2$ and $\epsilon=0.1$. 
Numerical simulation of
Eqs.~(2.4) and (2.5) was carried out for different values of
$K$. Figure 1a shows a typical case of large $K$ for which the spiral 
pattern rotates 
steadily and there is nothing anomalous.
\begin{figure}
\centerline{\includegraphics[width=14cm]
			            {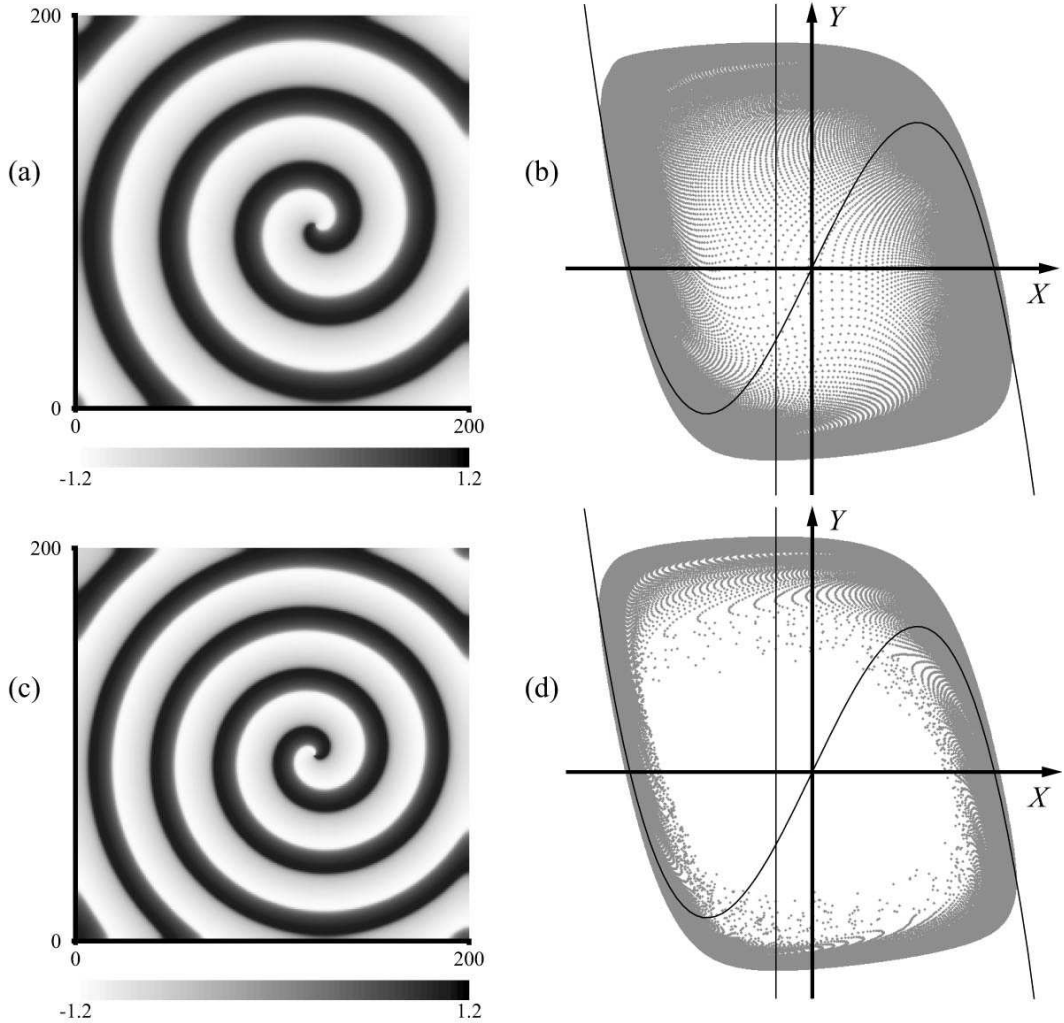}}
\caption{
Two-dimensional Spiral patterns for the component $X$ 
in gray scale
exhibited by the reaction-diffusion model
given by Eqs.~(2.1)$\sim$(2.3) in the oscillatory regime, 
and the corresponding phase portraits
projected on the $X$-$Y$ space. In the latter, the nullclines $f(Y,Y)=0$
and $g(X,Y)=0$ are also indicated. Parameter values are $K=10.0$ for (a)
and (b), and $K=4.0$ for (c) and (d); $a=1.0$, $b=0.2$ and $\epsilon=0.1$
are common. 
}
\label{fig1}	
\end{figure}
The corresponding phase portrait
at a given time
projected onto the $XY$ plane is shown in Fig.~1b. The phase portrait
is a set of points in the phase space representing
the current state of all local oscillators. 
In reaction-diffusion systems, we usually expect that this mapping 
between the physical space and the phase space is homeomorphic.
Hence, the inner region of the phase portrait
should be completely filled, or equivalently,
this object is simply connected in the continuum limit. Figure 1b confirms
this fact. 
The center of rotation
of the spiral is mapped to a certain fixed point 
$(X_0,Y_0, B_0)$ in the phase space. Under other parameter
conditions the spiral core may exhibit meandering. No such 
fixed point could exist then, still the phase portrait should remain simply 
connected. 

Let the coupling strength $K$ become smaller. 
The overall spiral pattern shown in Fig.~1c does not
seem very different from the strong coupling case except that the 
characteristic wavelength, which is comparable with 
the core radius, becomes smaller. However, the corresponding phase portrait
(see Fig.~1d) changes qualitatively. This object (called $\cal O$ hereafter) 
is no longer simply 
connected, and we clearly see
a large central hole. What does a hole mean physically? 
Imagine that we moved
along the frame of the spiral picture of Fig.~1c. The corresponding
trajectory in the phase space will trace the periphery of $\cal O$. 
We now let the square along which we move in the physical space
be shrunk a little. The corresponding phase trajectory will come a little 
inside the periphery of $\cal O$. Let the closed path in the physical space
be made smaller and smaller, down to an infinitesimal size. It is clear that, 
in the
presence of a hole in the phase portrait, there is a limitation beyond which
the phase trajectory can no longer be shrunk. This is apparently a 
contradiction as long as we adhere to a homeomorphism between the two
objects in the physical space and the phase space. We are forced to abandon
this property of the mapping, 
or we have to admit that a neighborhood in the physical space
can no longer be mapped to a neighborhood in the phase space. This implies
that the spatial continuity of the pattern has been lost 
which is likely to occur
in the central core.
If we look closely into the spiral core in the strong- and weak
coupling cases (see Fig.~2),
we actually find that there is nothing anomalous for the first case, while
for the second case a small group of oscillators near 
the core seem to be behaving individually rather than collectively.
\begin{figure}
\centerline{\includegraphics[width=14cm]
			            {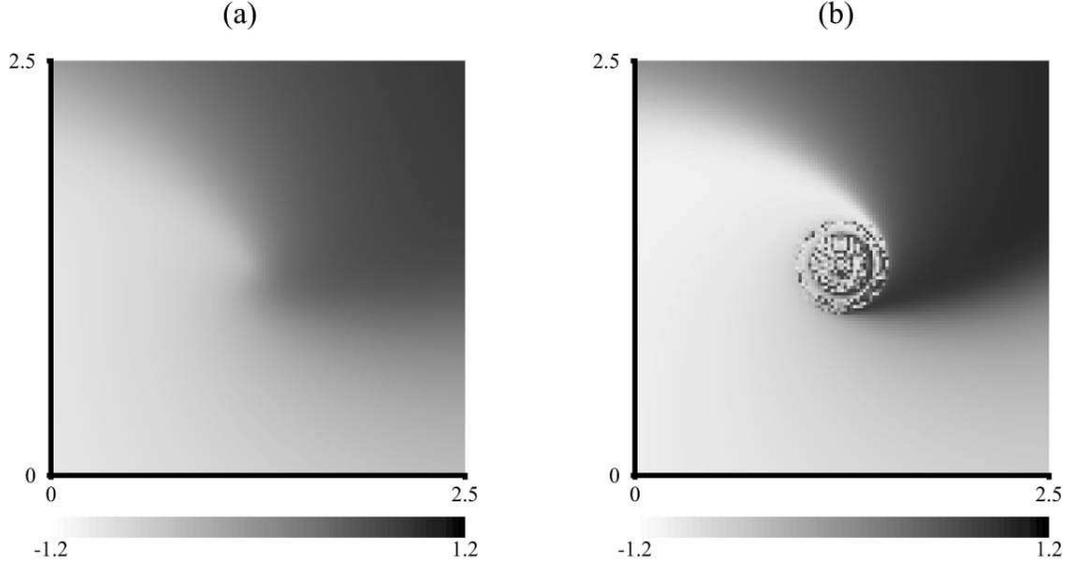}}
\caption{
Structures near the core of the spiral patterns in Fig.~1 (a) and
(c). $K=10.0$ for (a) and 4.0 for (b).
}
\label{fig2}	
\end{figure}
In the next section, we develop an efficient way of uncovering the origin
of this anomaly by means of a reduction of our evolution equations. 

\section{Case of non-locally coupled complex Ginzburg-Landau equation}
As a well-known fact, reaction-diffusion systems are generally reduced to
a complex Ginzburg-Landau equation near the Hopf bifurcation point of the local
oscillators\cite{rf:9}. The same is true of our reaction-diffusion system 
Eqs.~(2.1)$\simeq$(2.3) or their approximate form given by Eqs.~(2.4) and 
(2.5). Since spatial continuity can never be broken
in the complex Ginzburg-Landau equation, the
reduction method near the bifurcation is useless for our purposes.
As anticipated in the preceding section, the present type of 
anomaly arises as a result of effective non-locality in coupling, particularly
when the effective radius of coupling becomes comparable with the
characteristic scale of the pattern (typically the core radius). The failure
of the reduction method applied to the present problem comes from the loss
of effective non-locality near the bifurcation point. The reason is clear:
Near the bifurcation point, the characteristic wavelength of patterns becomes
as large as $\mu^{-1/2}$ in terms of the bifurcation parameter $\mu$, so that
the effective non-locality such as given by the last term in Eq.~(2.4)
disappears and can well be replaced with a diffusion term.

From the above reasoning, we may notice that there is still a way of
retaining effective non-locality near a bifurcation point if the coupling
$K$ is made as weak as $O(\mu)$\cite{rf:13}. Thus, applying an idea similar to the one
underlying the multiple bifurcation theory, we now try to reduce our
reaction-diffusion system near the codimension-two point 
$(\mu,K)=(0,0)$ by which effective non-locality may be recovered. 
The same idea was used previously in the study of multi-affine chemical
turbulence\cite{rf:6} for which effective non-locality was also crucial.

For the sake of simplicity, the third variable $B$ is assumed to change very 
rapidly, and we take the limit $\tau\rightarrow 0$. Thus, the set of equations
to be reduced is Eqs.~(2.4) and (2.5). As usual, the reduced equation is 
obtained in the form of an equation governing 
a complex variable $A(\bd{x},t)$. Under suitable rescaling of $A$
and the space-time coordinates, it takes the form
\begin{equation}
\dot{A}=(1+i\omega_{0})A-(1+ib)|A|^{2}A+K\cdot (1+ia)(Z-A),
\end{equation}
where
\begin{equation}
Z(\bd{x},t)=\int d\bd{x}'G(|\bd{x}-\bd{x}'|)A(\bd{x}').
\end{equation}
The coupling parameter $K$ in Eq.~(3.1) is not the original $K$ but the scaled 
$K$ with the factor $\mu^{-1}$, so that, by assumption, the latter $K$ has an
ordinary magnitude. We call Eq.~(3.1) 
non-locally coupled complex Ginzburg-Landau
equation (or simply the non-local CGL). It reduces to the ordinary CGL
when $K$ is sufficiently large, under the condition of which $A$ becomes 
so long-waved that the diffusion-coupling approximation of the last term
in Eq.~(3.1) would be valid.

For the purpose of gaining a qualitative understanding of how the solution
of Eq.~(3.1) behaves, it is sometimes useful to regard this equation as
describing a single oscillator driven by a forcing $Z$. Similarly, the
system as a whole may be regarded as an assembly of oscillators without
direct mutual coupling, but under the influence of a common forcing field
$Z$ which may depend on space as well as on time. The characteristic 
wavelength of $Z$ cannot be shorter than the decay length of $G$ which is
chosen to be 1. The forcing $Z$ is actually of {\em internal} rather
than external origin, 
and thus should be related to the totality of the individual
motion of the oscillators in a self-consistent manner. 
In the statistical-mechanical language, this may be called a mean-field 
picture which works exactly in the present system
because, in the continuum limit, infinitely many oscillators fall 
within the coupling range. 

\section{Core anomaly in the non-local CGL}
We want to know if a spiral-core anomaly similar to the one
observed for our reaction-diffusion model also arises in the non-local CGL,
especially when $K$ is small. 
It is well known that even the spiral waves in the ordinary CGL can 
exhibit complex behavior. Therefore, in order to separate the type of anomaly
to be caused by the non-locality in coupling from the conventional one,
we will work with in the parameter region in which the diffusion-coupling
approximation of Eq.~(3.1) admits a steadily rotating spiral solution.
Under the same condition, Eq.~(3.1) itself should exhibit steadily rotating
spiral waves provided $K$ is sufficiently large.     
It is easy to show that such a solution is bound to lose stability for 
sufficiently small $K$. The reason is the following. From the system's 
symmetry, the center of rotation of such a solution has a vanishing value
of $A$. From the same symmetry, $Z$ should also vanish there. 
This means that, if we work with the forced-oscillator picture mentioned
in \S 3, and apply it to the central oscillator,
then this oscillator is free of forcing and simply obeys the equation
\begin{equation}
\dot{A}=(1-K+i\omega_{0}')A-(1+ib)|A|^{2}A,
\end{equation}
where $\omega_{0}'=\omega_{0}-Ka$. 
It is clear that if $K>1$, vanishing value of $A$ is stable, while
if $K<1$ it is unstable contradicting the existence of steadily rotating
solution with vanishing central amplitude. Note that the above argument
does not guarantee the stability of the steadily rotating solution for
$K$ larger than 1.

Some results of our numerical simulation on Eq.~(3.1) for $K$ less than 1
are displayed in Fig.~3 to Fig.~5.
\begin{figure}
\centerline{\includegraphics[width=14cm]
			            {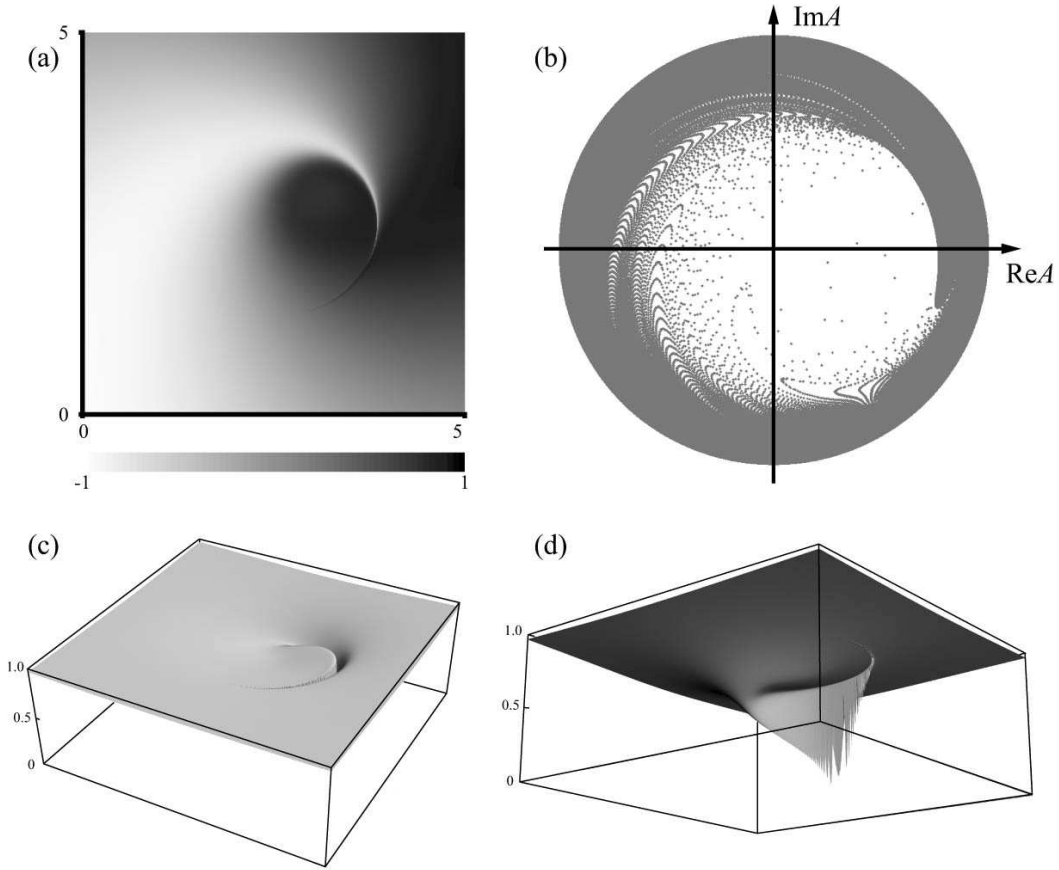}}
\caption{
Structure of the spiral core exhibited by the non-local
CGL given by Eq.~(3.1) with parameter values $K=0.5$, $a=0$ and $b=0.5$.
(a) Phase pattern in gray scale; (b) the corresponding phase portrait;
(c) and (d) 3D views of $|A|$ from the top and the bottom, respectively.  
}
\label{fig3}	
\end{figure}
\begin{figure}
\centerline{\includegraphics[width=6.5cm]
			            {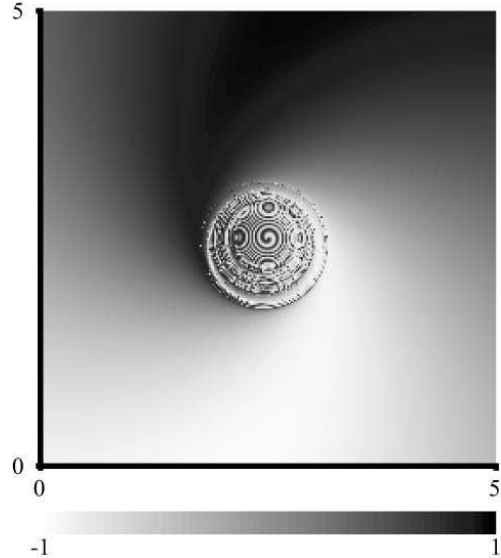}}
\caption{
Similar to Fig.3 (a) but for different parameter condition, i.e.,
$K=0.3$, $a=0$ and $b=0.4$.
}
\label{fig4}	
\end{figure}
\begin{figure}
\centerline{\includegraphics[width=14cm]
			            {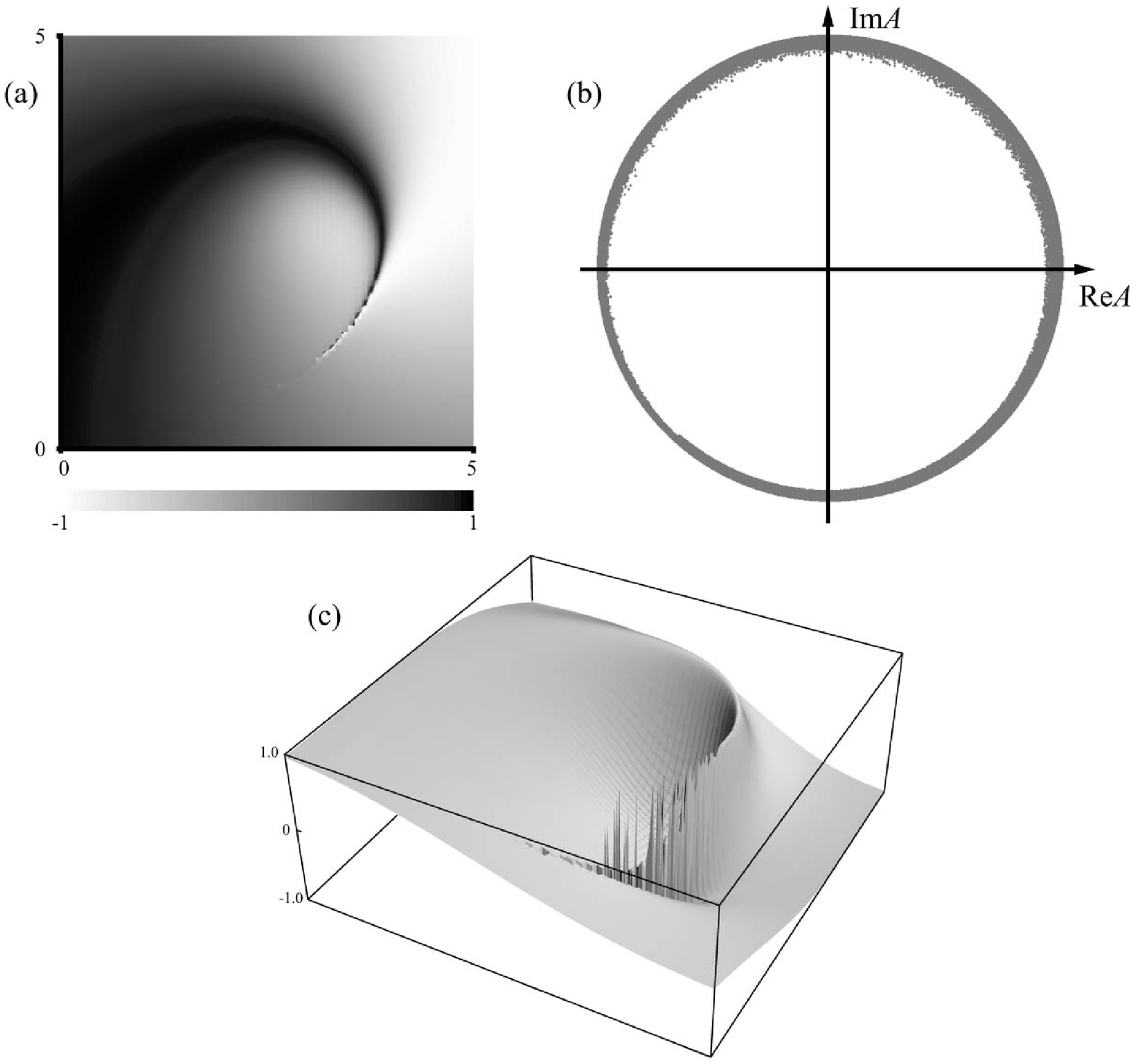}}
\caption{
(a) and (b) are similar to Fig.~3 (a) and (b), respectively,
but for much smaller value of $K$, i.e., $K=0.1$. (c) shows a 3D pattern of
Re~$A$.
}
\label{fig5}	
\end{figure}
A typical case is shown in Fig.~3 
from which a large hole is seen in the phase portrait.
The 3D views of the 
distribution of $|A|$ shows the appearance of a cut along which the 
oscillators 
behave incoherently with relatively small amplitude. The corresponding 
corrugated pattern never shows a sign of smoothening 
as the minimum spacing of the oscillators used in the numerical
simulation is made smaller.
This implies genuine spatial discontinuity distributed densely along the cut.
The cut may not be a true one-dimensional object, still this 
makes a sharp contrast
with the simulation on the original reaction-diffusion system displayed in
Fig.1 where the incoherence seems to have a two-dimensional extension.
The origin of such a difference may be understood from Fig.~4 showing a
simulation result under a different parameter condition. 
The incoherent structure in that figure
forms practically an extended object quite similar to Fig.~1.

It is interesting to observe that rotating spiral can persist when the coupling
$K$ is so small (as small as 0.1) that the phase portrait looks like a ring
leaving only a very thin periphery (see Fig.~5). 
This means that the amplitude degree of
freedom is almost dead. All oscillators are now oscillating with almost full
amplitude, i.e. $|A|\simeq 1$, so that the pattern of $|A|$ would not be 
interesting. Instead, a meaningful pattern is exhibited by the real part
of $A$ from which
the spatial discontinuity can well be seen. The absence of amplitude degree 
of freedom implies that the so-called phase oscillator model can well describe 
the same dynamics as the above, while our general view is that rotating 
spirals exhibited by phase oscillators are bound to lead to a topological 
contradiction\cite{rf:12}. The last statement does not apply, however, 
in the absence of spatial
continuity.

\section{Origin of spatial discontinuity and loss of smoothness}
In the preceding sections, we have contrasted the spiral dynamics between
the two cases of strong and weak couplings. In one case, the spiral core 
behaves normally while in the other case the dynamics is anomalous in the
sense that there is a group of oscillators inside the core which are
behaving individually. Although it would be interesting to know some more
details of how one type of behavior changes to the other as the coupling
constant $K$ is made smaller, numerical data showing a detailed scenario 
starting from a normal 
behavior changing to increasingly complex behavior are not available yet.
Still the following rough scenario seems to be correct at least.
Rigid rotation of the pattern first becomes unstable at some $K$ 
(say, $K=K_1$, which should definitely be less than 1), 
and successive bifurcations will occur up to another $K$ 
(say, $K=K_2$) above which the core becomes turbulent. In the turbulent
regime, the pattern is spatially continuous at first, but there is a third
critical value of $K$ (say, $K=K_3$) at which such continuity is lost and
the a group of oscillators inside the core lose mutual synchrony.

We are particularly interested in the situation near the transition 
when the synchrony is first lost and independent motion of the oscillators
sets in. From a theoretical point of view, there are good reasons to 
believe that over some finite range of $K$ preceding 
this transition, the pattern
loses spatial smoothness in a mathematical sense. The reason may most roughly
be stated as follows. As we pointed out in \S 2, something anomalous is expected
to occur when the coupling becomes so weak that the characteristic wavelength
 of the pattern which was denoted by $l_p$ becomes comparable or even smaller than the effective coupling 
radius. The quantity $l_p$ may be chosen as the core radius in our case.
This means that inside the core our system has no intrinsic length scale
down to the zero value as long as a continuum limit of oscillator distribution
is assumed. Thus, the inside region of the core is so to speak a {\em
scale-free world}. Once turbulent fluctuations occurred in this scale-free
world, they should also be scale-free. In other words, fluctuations should 
exist over all scales below the coupling radius 
down to infinitesimal scales in a self-similar manner, 
which implies loss of
spatial smoothness.

The above reasoning regarding the loss of spatial smoothness could be made
a little more mathematical as follows. One may ask first what physical 
quantities carry information on the smoothness and continuity of the pattern.   The most convenient quantities to work with seem to be various moments of
amplitude increment between two spatial points with infinitesimal mutual
distance. Let such amplitude increment over the distance $x$ 
be denoted by $y$, and its $q$th
moments by $<y^{q}>$. 
Detailed definition of ''amplitude'' is irrelevant. Real or imaginary part
of $A$ as well as $|A|$
may be chosen as $y$ if we are working with complex CGL, while $X$ or $Y$
(but not $B$) would be appropriate for $y$ for the reaction-diffusion system
given by Eqs.~(2.1)$\sim$(2.3). In what follows, we use terms appropriate for
the non-local CGL. 

The amplitude vs.\ $x$ curve which is fluctuating
may be regarded as continuous and differentiable to an arbitrary degree if
\begin{equation}
<y^q>\sim x^q 
\end{equation}  
in the limit $x\rightarrow 0$.
It was argued in previous works\cite{rf:5,rf:6,rf:10} 
that non-locally coupled oscillator systems
of the type we are now working with give moments $<y^q>$ different from
Eq.~(5.1). Remember that, as stated toward the end of \S 3, 
our system may be regarded as an assembly of 
independent oscillators driven by a common forcing. The forcing field
$Z(\bd{x},t)$ is temporally nonperiodic in general and its spatial variation 
should be smooth
with typical wavelength comparable with the coupling radius or core radius. 
Since the feedback 
from the motion of a given oscillator to the forcing field acting on it
is totally negligible, one may introduce a local Lyapunov exponent $\lambda(t)$
and its mean $\bar{\lambda}$ defined for the individual oscillator.
Well inside the core, $\lambda(t)$ and $\bar{\lambda}$ are expected to be
space-independent. Thus, regarding possible statistics of $<y^q>$, one may
apply a previous theory on multi-affine spatio-temporal chaos in non-locally
coupled oscillatory fields. The theory tells that the behavior of $<y^q>$
depends crucially on the sign of $\bar{\lambda}$. If $\bar{\lambda}<0$, it
is given by
\begin{equation}
<y^q>\sim x^{\zeta(q)}
\end{equation}
with some nonlinear function of $q$, while if $\bar{\lambda}>0$, we have
\begin{equation}
<y^q>\sim \delta+x^{\zeta(q)},
\end{equation}
where $\delta$ is a positive constant signifying spatial discontinuity. 
If $\bar{\lambda}$ is negative for a given oscillator, its motion
should be synchronized
with the forcing field, so that a small group of oscillators composed of
its neighbors will also be mutually synchronized, thus leading to
spatial continuity. If $\bar{\lambda}$ becomes positive, such mutual synchrony
is lost and individual motion of the oscillators sets in.
This type of transition is expected to occur 
at a certain critical value of $K$.

It can further be shown that the exponent $\zeta(q)$ is identical with $q$ 
up to some value $q=\beta$, while it completely saturates to a constant, i.e.,
$\zeta(q)=\beta$ for $q>\beta$. Thus, prior to the onset of discontinuity,
the pattern loses complete smoothness in the sense that the moments of $y$
higher than the order $\beta$ behaves anomalously. $\beta$ is shown to
tend to zero as the onset of discontinuity is approached, by which all moments
become anomalous. It should also be remarked that the pattern becomes fractal
if $\beta<1$. 

In a previous work\cite{rf:11}, 
these theoretical results were confirmed
by numerical simulation for systems with statistical homogeneity in space.
For the present type spiral-core anomaly, in contrast, the turbulent regime
is strongly localized, so that such comparison between the theory and 
numerical experiment would not be easy.
\section{Concluding remarks}
We have shown that in non-locally coupled systems 
spiral core is the weakest part
of rotating spiral waves in the sense that
localized turbulence without spatial smoothness and even spatial continuity
is initiated there. We have also shown that such behavior is not
restricted to systems of genuine non-local coupling but may also arise
in an important class of reaction-diffusion systems where some of the chemical
components are diffusion-free. The non-local CGL obtained near the 
codimension-two point proved to be extremely useful 
for the study of the dynamics
peculiar to non-locally coupled self-oscillatory fields. 

Complete absence of local coupling on which the whole argument of the
present paper relies would not be very
realistic in real reaction-diffusion systems. How the spiral dynamics 
changes when weak diffusive coupling has been introduced 
could be an interesting
future problem.


\begin{thebibliography}{99}
\bibitem{rf:1} D.~Barkeley, 
  in {\em Chemical Waves and Patterns}, ed.~R.~Kapral and 
   K.~Showalter, Dordrecht, Kluwer, Ch.~5 p.~163 (1995).
\bibitem{rf:2} E.~Mihaliuk et al., Faraday Discuss., {\bf 120}, 383 (2001).
\bibitem{rf:3} J.~P.~Keener, Physica D {\bf 31}, 269 (1988).
\bibitem{rf:4} A preliminary results of this work were reported in: 
  Y.~Kuramoto, in {\em Nonlinear Dynamics and Chaos: Where do we go   from here?},   ed.~S.~J.~Hogan et al., IoP, p.~209 (2003).
\bibitem{rf:5} Y.~Kuramoto, Prog.~Thor.~Phys. {\bf 94}, 321 (1995).
\bibitem{rf:6} Y.~Kuramoto, D.~Battogtokh and H.~Nakao, 
Phys.~Rev.~Lett. {\bf 81}, 
  3543 (1998).
\bibitem{rf:8} Y.~Kuramoto, H.~Nakao and D.~Battogtokh, 
   Physica A {\bf 288}, 244 (2000).
\bibitem{rf:7} V.~K.~Vanag and I.~R.~Epstein, Phys.~Rev.~Lett. {\bf 88}, 
088303 (2002).
\bibitem{rf:9} Y.~Kuramoto, {\em Chemical Oscillations, Waves, and Turbulence},    Springer-Verlag,(1984).
\bibitem{rf:13} D.~Tanaka and Y.~Kuramoto, in preparation.
\bibitem{rf:12} A.~T.~Winfree, {\em The Geometry of Biological Time} (2nd Ed.),
  Springer, (2000).
\bibitem{rf:10} Y.~Kuramoto and H.~Nakao, Phys.~Rev.~Lett. 
{\bf 76}, 4352 (1996).
\bibitem{rf:11} H.~Nakao and Y.~Kuramoto, Eur.~Phys.~J. B{\bf 11}, 345 (1999).
\end{thebibliography}
\end{document}